\documentclass[aps,prd,showpacs,showkeys,superscriptaddress,twocolumn]{revtex4}
\usepackage{epsfig}
\usepackage{graphicx}
\usepackage{amsmath,amssymb,amsfonts,latexsym}
\usepackage{graphicx}

\usepackage{epsfig,amssymb,amsfonts,verbatim}

\usepackage{amsmath}
\usepackage{latexsym}
\usepackage{amsfonts}
\usepackage{amssymb}
\usepackage{color}

\def\bfl{\begin{flushleft}}
\def\efl{\end{flushleft}}
\def\bfr{\begin{flushright}}
\def\efr{\end{flushright}}
\def\bc{\begin{center}}
\def\ec{\end{center}}

\def\ba{\begin{eqnarray}}
\def\ea{\end{eqnarray}}
\def\baa#1{\begin{array}{#1}}
\def\eaa{\end{array}}
\def\bw{\begin{widetext}}
\def\ew{\end{widetext}}
\def\nn{\nonumber }

\def\text#1{\mbox{#1}}

\begin{document}

\title{Length-dependent resistance model for a single-wall Carbon nanotube}

\author{Andrew Das Arulsamy}
\email{andrew@physics.usyd.edu.au}

\author{Marco Fronzi}
\address{School of Physics, The University of Sydney, Sydney, New South Wales 2006, Australia}

\date{\today}

\begin{abstract}
The non-linear length-dependent resistance, $\mathcal{R}(l)$
observed in single-wall Carbon nanotubes (SNTs) is explained
through the recently proposed ionization energy ($E_I$) based
Fermi-Dirac statistics ($i$FDS). The length here corresponds to the
Carbon atoms number ($\mathcal{N}$) along the SNT. It is also
shown that $\mathcal{R}_y(l_y)$ $<$ $\mathcal{R}_x(l_x)$ is
associated with $E_I^y$ $<$
$E_I^x$, which can be attributed to different conducting properties in their respective
$y$ and $x$ directions, or due to chirality.
\end{abstract}


\maketitle

\section{Introduction}

Enormous amount of research have been poured since the discovery
of Carbon (\texttt{C}) nanotubes (CNTs) by Iijima~\cite{iijima1} in 1991 and
consequently, CNTs have been successfully exploited to produce cathode ray tubes~\cite{mann2} and nano-electronic devices~\cite{vasili}. Understandably, CNTs are
believed to pave the pioneering pace for the nanotechnology boom.
Basically, \texttt{C} can be categorized into graphite, diamond and
Fullerenes based on their bonding nature that gives rise to
different electronic and structural properties. Unexpectedly,
\texttt{C} in all these three structures with slight manipulations
have exposed superconductivity~\cite{baskaran8,baskaran9,perfetto}. CNTs' electronic properties are
equivalent to rolled-graphite~\cite{dresselhaus3,lai,li,wu} which also
reveal superconductivity in the absence of
doping~\cite{sheng4,kociak5,matsudaira} and concentration-dependent
non-linear optical properties. The real part of third-order
non-linear susceptibility, Re $\chi^{(3)}$ was found to be in the
order of 10$^{-11}$ esu for multi-wall CNTs by Elim's
group~\cite{elim6}. This value is roughly 100$\times$ larger than
that of SNTs, which is due to SNT's lower \texttt{C}-atom
concentration. The superconducting properties of Boron-doped
diamonds~\cite{ekimov7} based on resonating-valence-bond mechanism
was put forward by Baskaran~\cite{baskaran8,baskaran9} whereas,
the superconducting Fullerenes and its non-linear optical
properties have been discussed by Cohen {\it et
al.}~\cite{cohen10} and Elim {\it et al.}~\cite{elim11}
respectively.

Here, the ionization energy based Fermi-Dirac statistics (iFDS) is
employed to derive the length-dependent resistance model,
$\mathcal{R}(l)$. The derivation of iFDS and its
applications in a wide variety of strongly correlated electronic
matter is given in the
Refs.~\cite{andrew25,andrew26,andrew27,andrew28}.
This model is shown to be viable in addressing the recent
$\mathcal{R}(l)$ observation reported by de Pablo {\it et
al.}~\cite{pablo33}, Andriotis {\it et al.}~\cite{andriotis34} and Purewal {\it et al.}~\cite{purewal}
in CNTs. The length-dependent resistance is an intrinsic property
basically because the contact resistance is independent of CNT's
length~\cite{pablo33}. As a consequence, the only questionable
result is the magnitude of the resistance, not its
length-dependent trend. However, other measurements namely, the
temperature($T$)-dependent electrical or heat conductance are strongly
influenced by the contact resistance due to its own $T$-dependence
and its large magnitude, usually in the order of the CNTs
resistance, which in turn waver the intrinsic experimental
$\mathcal{R}(T)$ results. It is interesting to note that the resistance of a SNT is non-linearly proportional to the tube's length in both metallic and semiconducting SNTs~\cite{purewal,zhang,suzuura} at any given $T$. However, the calculations carried out by Zhang {\it et al.}~\cite{zhang} and Uryu {\it et al.}~\cite{uryu1,uryu2} for metallic CNTs indicate that the resistance is inversely proportional to the length as a result of resonant tunelling at interface. In this work, we do not consider heterostructures with resonant tunneling, but rather, on intrinsic metallic and semiconducting SNTs. The resistance model derived here are also suitable in other strongly correlated nanotubes that allow direct-current resistance and/or polarization measurements, or if the \texttt{C} atoms in CNTs are doped substitutionally with different atoms.

\section{The length-dependent resistance model}

We start with the many-body Hamiltonian~\cite{andrew1,andrew2}, 

\begin {eqnarray}
-\frac{\hbar^2}{2m}\nabla^2\varphi = (E + V(\textbf{r}))\varphi, \label{eq:1}
\end {eqnarray}

of which, 

\begin {eqnarray}
\hat{H}\varphi = (E_0 \pm \xi)\varphi. \label{eq:1002}
\end {eqnarray}

From Eq.~(\ref{eq:1002}), one can notice that the
influence of the potential energy on the total energy has been conveniently parameterized as $\xi$.
This energy function, $\xi$ can be characterized in such a
way that $E_0$ is the total energy, $E$ at $T$ = 0. Add to that, from Eq.~(\ref{eq:1002}), it
is obvious that the magnitude of $\xi$ is given by $\xi = E_{\rm{kin}} - E_0 + V(\textbf{r})$, $E_{\rm{kin}}$ denotes the kinetic energy. Physically, $\xi$ implies the energy needed to overcome
the potential energy that exists in a particular system. That is, $\xi$
is the energy needed to excite a particular electron to a finite distance, $r$, not necessarily $r \rightarrow \infty$. Literally, this is exactly what we need to know in any condensed matter that actually or reasonably defines the
properties of the fermions. $\hat{H}$ is the usual Hamilton operator, $\varphi$ denotes the many-body eigenstate and $E_0$ is the total energy at $T$ = 0. The + sign of $\pm\xi$
is for the electron ($0 \rightarrow +\infty$) while the $-$ sign
is for the hole ($-\infty \rightarrow 0$). In addition, we define the ionization energy in a many-atom system, $\xi = E_I^{\rm{real}}$ is approximately proportional to $E_I$ of an isolated atom or ion. We can prove the validity of Eq.~(\ref{eq:1002}) by means of constructive (existence) and/or direct proofs as given in Ref.~\cite{andrew2}. However, for an isolated atom, $\xi$ is given by 

\begin {eqnarray}
\pm \xi = E_{\rm{kin}} - E_0 + V_{\rm{pot}} = \pm E_I, \label{eq:a1}
\end {eqnarray}

The corresponding total energy is 

\begin {eqnarray}
&E_0 \pm \xi &= E_{\rm{kin}} + V_{\rm{pot}} \nn \\&& = E_0 \pm E_I. \label{eq:a2} 
\end {eqnarray}

On the other hand, for an atom in a many-atom system, we can rewrite Eq.~(\ref{eq:a1}) as

\begin {eqnarray}
&\pm \xi &= E_{\rm{kin}} - E_0 + V_{\rm{pot}} + V_{\rm{many-body}} \nn \\&& = E_I
+ V_{\rm{many-body}} \nn \\&& = \pm E^{\rm{real}}_I. \label{eq:a3} 
\end {eqnarray}

Note here that $V_{\rm{pot}}$ is the atomic Coulomb potential, while the $V_{\rm{many-body}}$ is the many body potential averaged from the periodic potential of the atomic arrangement. The corresponding total energy from Eq.~(\ref{eq:a3}) is given by 

\begin {eqnarray}
&&E_0 \pm \xi = E_{\rm{kin}} + V_{\rm{pot}} + V_{\rm{many-body}} \nn \\&& = E_0
\pm E_I + V_{\rm{many-body}} \nn \\&& = E_0 \pm E^{\rm{real}}_I. \label{eq:a4} 
\end {eqnarray}

In this case, $E_I^{\rm{real}}$ is the ionization
energy of an atom in a many-atom system (not isolated). The exact values of $E_I$ are known
for an isolated atom. As a consequence, we can arrive at Eq.~(\ref{eq:1002}) from Eq.~(\ref{eq:a4}). Apparently, we cannot use Eq.~(\ref{eq:1002}) to isolate the electronic and phonon contributions because we have defined the $\xi$ as a function of the Coulomb potential ($V_{\rm{pot}}$), many-body ($V_{\rm{many-body}}$) and kinetic ($E_{\rm{kin}}$) energies. Consequently, the total energy can also be rewritten as (from Eq.~(\ref{eq:a4}))  

\begin {eqnarray}
E = E_0 \pm \sum_i^z\sum_j E_{Ii,j}^{\rm{real}}, \label{eq:a5}
\end {eqnarray}

where, $j$ is the sum over the constituent elements in a particular compound. For a \texttt{C} nanotube with only one type of atom, Eq.~(\ref{eq:a5}) can be rewritten as

\begin {eqnarray}
E = E_0 \pm \beta \sum_i^z E_{Ii}. \label{eq:a6}
\end {eqnarray}

In Eq.~(\ref{eq:a6}), we have defined here that $\beta = 1 + \frac{\langle V(\textbf{r})\rangle}{E_I}$, where $\langle V(\textbf{r})\rangle$ is the averaged many-body potential value. Apart from that, the total energy equation for a free-electron system is given by  

\begin {eqnarray}
&E& = E_0 \pm \sum_i^z\sum_j E_{Ii,j}^{\rm{real}} \nn \\&& = E_0 \pm
[E_{\rm{kin}} - E_0 + V_{\rm{pot}} + V_{\rm{many-body}}] \nn \\&& = E_{\rm{kin}} +
V_{\rm{pot}} + V_{\rm{many-body}} \Leftrightarrow ~\rm{for~
electrons}~ \pm \rightarrow + \nn \\&& = E_{\rm{kin}} + V_{\rm{total}} \nn \\&& = E_{\rm{kin}} \Leftrightarrow \rm{implies~free ~
electrons}. \label{eq:a7} 
\end {eqnarray}

In Eq.~(\ref{eq:a7}) we have substituted Eq.~(\ref{eq:a3}) for $E_I^{\rm{real}}$ because the
concept of ionization energy is irrelevant for free-electron metals, which do not require excitations
from its parent atom to conduct electricity. As such, the carrier density is independent of temperature and the
scattering rate is the one that determines the resistivity with
respect to temperature, impurities, defects, electron-electron and
electron-phonon interactions. In summary, the total energy from Eq.~(\ref{eq:1002}) carries the
\textit{fingerprint} of each \texttt{C} atom in a nanotube and it refers to the difference in the
energy levels of each atom rather than the absolute values of each
energy level (eigenvalues) in each atom. Hence, the kinetic energy of
each electron from each atom will be captured by the total energy and preserves the atomic level
\textit{electronic-fingerprint} in the nanotube. Using this newly defined total energy, we can derive the ionization energy based Fermi-Dirac statistics ($i$FDS) as given below~\cite{andrew25} 

\begin{eqnarray}
&&f_e(E_0,\xi) = \frac{1}{e^{[\left(E_0 + \xi
\right) - E_F^{0}]/k_BT }+1}, \nn
\\&& f_h(E_0,\xi) = \frac{1}{e^{[E_F^{0} - \left(E_0 - \xi
\right)]/k_BT}+1}. \label{eq:1003}
\end{eqnarray}

where, $E_F^{0}$ is the Fermi level at $T$ = 0 and $k_B$ is the Boltzmann constant. However, substituting the same atom in a nanotube gives rise to
the influence of many-body $V(\textbf{r})$ and in reality, $E_I^{real}$ cannot be
evaluated from Eq.~(\ref{eq:a3}). Nevertheless, the $E_I^{real}$
of an atom or ion in a nanotube is proportional to the isolated
atom and/or ion's $E_I$ as given in Eq.~(\ref{eq:a6}). It is this property that enables one to predict the variation of fermionic excitation probability in \texttt{C} nanotubes. Therefore, one can employ the experimental atomic spectra to
estimate, $\xi$ = $E_I^{real}$ $\propto$ $E_I$. It is emphasized here that $E_I$ is zero for
Boltzmann particles. As such, one should not assume that the above
approximation should give the Boltzmann distribution function
(BDF) as a classical limit. One can indeed arrive at BDF by first
denying the additional constraint by substituting $E_I$ = 0. Importantly, Eq.~(\ref{eq:1003}) is the Fermi-Dirac statistics derived specifically for strongly correlated matter, where it is \textit{not} applicable for free-electron system (or Fermi gas) as shown in Eq.~(\ref{eq:a7}). 

Now, before we move on, let us re-examine Eq.~(\ref{eq:a3}) that seems to say nothing about i) the atomic arrangements and ii) how to isolate the phonon from electronic contribution. Firstly, Eq.~(\ref{eq:a3}) is perfectly applicable for any atomic arrangements or crystal structures. The reason is that we can incorporate Eqs.~(\ref{eq:a3}) and~(\ref{eq:1003}) for both non-bulk system, namely SNTs as well as for bulk system, regardless of its specific crystal structures, since these two equations can be normalized by employing the appropriate density-of-states (DOS). However, for non-bulk system of several atoms, including SNTs, we need to incorporate the atomic arrangement explicitly because the electronic excitation depends on the number of atoms along a certain conducting path (developed here). For bulk system with the number of atoms of the order of 10$^{23}$, the effect of different crystal structures do not arise because the conducting paths are isotropic and the $E_I$ here will and can be dressed accordingly to take this structural effect into account~\cite{andrew25,andrew26,andrew27,andrew28,andrew1}. For example, pure diamond and graphite will each have different valence states and electronic polarizabilities (the ability of the valence electrons to excite in a particular direction in the presence of electric field), in which, these differences are due to the different excitations of the valence electrons. These different excitations of the valence electrons are the ones that have been captured by Eq.~(\ref{eq:1003}) through Eq.~(\ref{eq:a3}). Therefore, in our approach, the true DOS and/or atomic arrangements of a particular system are unnecessary. The price we pay for this is that we cannot calculate the many-body eigenstates from Eq.~(\ref{eq:1002}), but note here that we \textit{can} indeed prove Eq.~(\ref{eq:1002}) microscopically for real isolated atoms~\cite{andrew2}. In other words, our input parameter is the isolated atomic-energy-level-difference, or defined here as the ionization energy ($E_I$). The theoretical discussion of how $E_I$ affect the polarizability can be found in Ref.~\cite{andrew28}. 

The second issue here is how do we isolate the phonon from the electronic counterpart? Basically, we cannot and there is no reason to, at least for condensed matter that violate free-electron metals and for as long as we do not apply this formalism to evaluate thermal conductivity. The next question is, how $E_I$ is related to electron-phonon interaction in the first place? We will answer this shortly. The 1-dimensional (1D) DOS is given by $N_e(E$,1D) =
$\big[E^{-1/2}(m^*_e/2)^{1/2}\big]/\pi\hbar$, using $E$ =
$\hbar^2k^2/2m^*_e$ and $k$ denotes the wave vector. The integral
to compute carrier density and its solution are given by (after making use of Eq.~(\ref{eq:1003})) 

\begin{eqnarray}
&n& = \int\limits^{\infty}_0{f_e(E_0,E_I)N_e(E)dE} \nn
\\&& =
\bigg[\frac{k_BTm^*_e}{2\pi\hbar^2}\bigg]^{1/2}\exp\bigg[\frac{E_F^0-E_I}{k_BT}\bigg].
\label{eq:1}
\end{eqnarray}

Based on Eq.~(\ref{eq:1}), suppose that the system is at temperature $T$ and it has $n$ number of electrons per unit volume. Now, imagine that we reduce the magnitude of $E_I$ (small enough that it does not increase $n$), then the only parameter that can change is the effective mass of the electron, where $m^*_e \propto E_I$, which in turn implies that the electron-phonon coupling ($\lambda_{\rm{el:ph}}$) has been reduced. This same argument with small $E_I$ variations can be applied at any reasonable temperatures. However, for free electron metals, $\lambda_{\rm{el:ph}}$ is defined as the electron-phonon scattering, where electrons and phonons can be treated as two different entities that scatter each other. In our approach, we do not treat the electrons and phonons, even in metallic SNTs as separate entities. In addition, $E_I$ in this case has no relation with electron-phonon scattering. Switching back to the SNT, the charge ($q$)-gradient along a nanotube's length ($l$) and the drift velocity ($v_d$) of charges can be written as~\cite{halliday35}

\begin{eqnarray}
\frac{dq}{dl} = n\pi de;~~~\frac{dl}{d\tau} = v_d. \label{eq:2}
\end{eqnarray}

As such, one can write the current ($i$) as~\cite{halliday35}

\begin{eqnarray}
i = \frac{dq}{d\tau} = \frac{dq}{dl} \times \frac{dl}{d\tau} = n\pi dev_d. \label{eq:3}
\end{eqnarray}

Now, the resistance for a single conducting path or length, of a
SNT is 

\begin{eqnarray}
\frac{\mathcal{R}}{\pi d} = \frac{V}{i} = \frac{1}{i}\int_0^{\pi d} \textbf{E}~ dr, \label{eq:c1} 
\end{eqnarray}

$d$ denotes the tube's diameter and $\textbf{E}$
= electric field. We also know that $m(dv_d/d\tau) = -e\textbf{E}$
that eventually gives $v_d = -e\textbf{E}\tau/m$. Finally, one can
arrive at the resistance of a whole nanotube, as given below

\begin{eqnarray}
&&\mathcal{R}(l) = \frac{(\pi d)^2 \textbf{E}}{i} = \pi
d\frac{m}{ne^2\tau} = \pi d \rho(E_I)\nn
\\&& = \pi d\frac{A\hbar}{e^2}\bigg(\frac{2\pi m^*_e}{k_B}\bigg)^{1/2}
T^{3/2}\exp\bigg[\frac{E_I-E_F^0}{k_BT}\bigg] \nn
\\&& = \pi d A(13062)\exp\bigg[\frac{E_I-E_F^0}{k_BT}\bigg]. \label{eq:4}
\end{eqnarray}

We have substituted, $m/ne^2\tau_{e-e}$ for $\rho (E_I)$ and the
electron-electron scattering rate, 1/$\tau_{e-e}$ = $AT^2$.
The $\tau_{el:ph}$ has been neglected because SNTs are not free-electron metals, even the metallic ones. However, for heat transport, $\tau_{el:ph}$ is not negligible. $A$ is the $T$-independent scattering rate
constant. The numerical value is obtained for $T$ = 300 K. The 1D resistivity, $\rho (E_I)$ for nanotubes can be written as 

\begin{eqnarray}
\rho(E_I) = \frac{A\hbar}{e^2}\bigg(\frac{2\pi m}{k_B}\bigg)^{1/2}T^{3/2}\exp\bigg[\frac{E_I - E_F^0}{k_BT}\bigg]. \label{eq:4a}
\end{eqnarray}

\begin{figure}[hbtp!]
\begin{center}
\scalebox{0.5}{\includegraphics{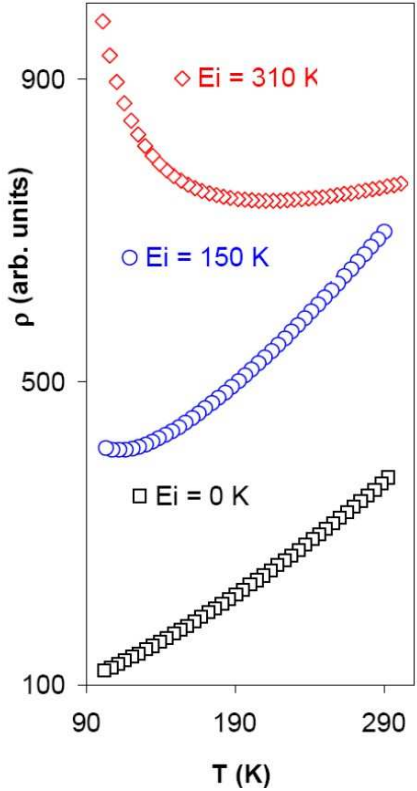}}
\caption{The intrinsic $T$-dependence of 1D resistivity is $T^{3/2}$ for $T$ above $E_I$. For $T$ below $E_I$, $\rho(E_I)$ is proportional to $\exp(1/T)$. There are three different curves for different magnitudes of $E_I$.}  
\label{fig:1}
\end{center}
\end{figure}

The calculated curves from Eq.~(\ref{eq:4a}) are shown in Fig.~\ref{fig:1}. Interestingly, one of the curve ($E_I$ = 150 K) is comparable with the experimental data in Ref.~\cite{purewal} (see Figure 2b). The calculation of the total average ionization energy (in the respective $y$ and $x$ directions) can be carried out with

\begin{eqnarray}
E_I^{y,x} [\texttt{C}^{z+}] =
\sum_i^z\sum_j\frac{E_{Ii}^{y,x}}{z}\mathcal{N}_j. \label{eq:5}
\end{eqnarray}

Unlike ionic bulk systems, CNTs are 1D systems with anisotropic conducting paths, which have covalent bonds. Consequently, the following definitions and descriptions are essential. The
$\texttt{C}$ in Eq.~(\ref{eq:5}) represents the Carbon atom while
$z$ denotes the number of valence electrons that can be excited, which will eventually contributes to the conductance
of CNTs in the presence of applied voltage. Apart from
$\mathcal{N}$ (the number of \texttt{C} atoms along a conducting path), the number of valence electron that are excited for conduction in the $y$
direction is not equal to the $x$. Meaning, the strength of the resistance or conductance in their respective $y$ and $x$
directions of a SNT originate from the inequality,
$E_I^y(\mathcal{N})$ $<$ $E_I^x(\mathcal{N})$. The subscripts, $i$ = 1, 2, ... $z$ and $j$ adds the \texttt{C} atoms,
1, 2, and so on continuously along its conducting path or length.
In the previous work on
superconductors and ferroelectrics~\cite{andrew25,andrew26,andrew27,andrew28}, Eq.~(\ref{eq:5}) was simply written as the average ionization
energy of a single ion as given in 

\begin{eqnarray}
E_I = \sum_i^z \frac{E_{Ii}}{z}. \label{eq:aa6}
\end{eqnarray}

The relative magnitude of $E_I$ was
then calculated based on the percentage of dopant to predict the
variation of $\rho(T)$ and dielectric constant. On the contrary,
SNTs with finite length in nanoscale and with only one type of atoms namely, \texttt{C} requires $E_I$ in
the form of Eq.~(\ref{eq:5}). 

\begin{figure}[hbtp!]
\begin{center}
\scalebox{0.65}{\includegraphics{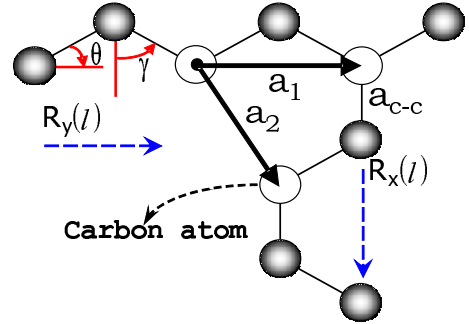}}
\caption{The arrangement of \texttt{C} atoms in a single-wall Carbon
nanotube is shown schematically. The resistance is strongly
influenced by the direction ($y,x$) in which $\mathcal{R}_{y,x}$
is measured. Such observation is due to the inequality,
$E_I^y(\mathcal{N})$ $<$
$E_I^x(\mathcal{N})$. The defined angles,
$\theta$ and $\gamma$ can be used to compute the length as given in
the Eqs.~(\ref{eq:6}) and~(\ref{eq:7}), respectively. $a_{c-c}$
denotes the length of covalent bond between two \texttt{C} atoms.}   
\label{fig:2}
\end{center}
\end{figure}

Figure~\ref{fig:2} schematically
shows the arrangement of \texttt{C} atoms in the $y$ and $x$
directions. Taking $a_{c-c}$ (0.142 nm) as the distance between
the two \texttt{C} atoms, one can write

\begin{eqnarray}
\mathbb{L}_y = \mathbb{L}_x = a_{c-c}\sum_{j=2}\mathcal{N}_j-1.
\label{eq:6}
\end{eqnarray}

The $\mathbb{L}_y$ and $\mathbb{L}_x$ are the lengths along the \texttt{C}$-$\texttt{C} atom's bond. Therefore, the experimentally measurable lengths (in real space) can be written as (as a function of $\mathbb{L}$) 

\begin{eqnarray}
l_y = \mathbb{L}_y\cos(\theta) =
a_{c-c}\cos(\theta)\sum_{j=2}\mathcal{N}_j-1.\label{eq:7}
\end{eqnarray}

\begin{eqnarray}
&l_x& = \frac{1+2\cos(\gamma)}{3}\mathbb{L}_x \nn\\&& =
\frac{a_{c-c}}{3}\big[1+2\cos(\gamma)\big]\sum_{j=2}\mathcal{N}_j-1.\label{eq:8}
\end{eqnarray}

Recall here that the reason $l$ in Eqs.~(\ref{eq:7}) and~(\ref{eq:8}) are written as functions of $\mathbb{L}$ is to take into account the higher probability of electrons to conduct along the $\mathbb{L}$. The angles, $\theta$ is the chiral angle, while $\gamma$ = 90$^{\rm{o}} - \theta$, which are also defined in Fig.~\ref{fig:2}. The subscript, $j$ = 2 indicates the sum starts
from the second \texttt{C} atom and so on. The chiral vector, $C_h$ is given by~\cite{odom} 

\begin{eqnarray}
C_h = n\textbf{a}_1 + m\textbf{a}_2, \nn
\end{eqnarray}

where $\textbf{a}_1$ and $\textbf{a}_2$ denote the 2D graphene lattice vectors, while $n$ and $m$ are integers. $C_h$ can be related to $x$($\theta$ = 0$^{\rm{o}}$) and $y$($\theta$ = 30$^{\rm{o}}$) with

\begin{eqnarray}
&&C_h(x) = n\textbf{a}_1 + n\textbf{a}_2, \nn \\&&
C_h(y) = n\textbf{a}_1. \nn
\end{eqnarray}

Consequently, Eq.~(\ref{eq:5}) in $y$ and $x$ directions can be respectively
rewritten as

\begin{eqnarray}
E_I^{y} [\texttt{C}^{z+}] =
\sum_i^z\frac{E_{Ii}^{y}}{z}\bigg(\frac{l_y}{a_{c-c}\cos(\theta)}+1\bigg).
\label{eq:9}
\end{eqnarray}

\begin{eqnarray}
E_I^{x} [\texttt{C}^{z+}] =
\sum_i^z\frac{E_{Ii}^{x}}{z}\bigg(\frac{3l_x}{a_{c-c}\big[1+2\cos(\gamma)\big]}+1\bigg).
\label{eq:10}
\end{eqnarray}

Now, one can actually substitutes either Eq.~(\ref{eq:9}) or
Eq.~(\ref{eq:10}) accordingly into Eq.~(\ref{eq:4}) in order to
obtain the length-dependent resistance. In addition, we can see that both Eqs.~(\ref{eq:9}) and~(\ref{eq:10}) are also determined by the chiral vectors.  

\section{Analysis of $\mathcal{R}(l)$}

The $\mathcal{R}(l)$ of free-electron metals with isotropic
distribution of atoms and electrons can be simply derived as
$\mathcal{R}(l) = \frac{\rho l}{S}$, $S$ denotes the cross section
area~\cite{halliday35}. However, CNTs resistance at 300 K, say in
the $x$ direction should be written as

\begin{eqnarray}
&\mathcal{R}_x& = \pi d\rho(E_I) \nn
\\&& = \pi dA(13062) \nn
\\&& \times
\exp\bigg\{\bigg[\bigg(\frac{3l_x}{a_{c-c}\big[1+2\cos(\gamma)\big]}+1\bigg)E_I-E_F^0\bigg]\frac{1}{k_BT}\bigg\}
\nn
\\&& \approx \pi dA(13062)\exp\bigg\{\frac{Bl_x}{T}\bigg\}.\label{eq:11}
\end{eqnarray}

\begin{figure}[hbtp!]
\begin{center}
\scalebox{0.55}{\includegraphics{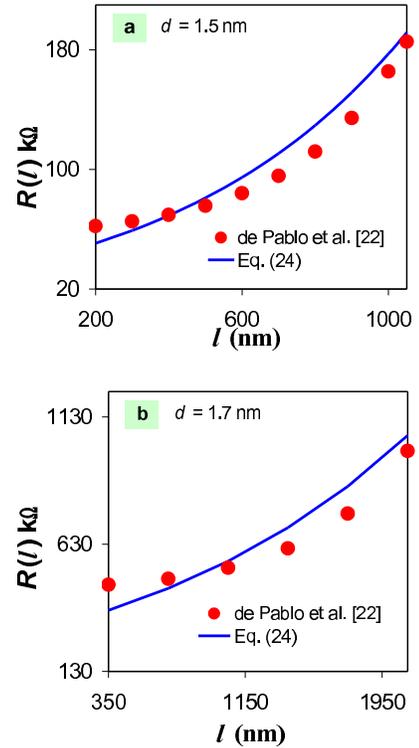}}
\caption{The length-dependent resistance ($\mathcal{R}$) based on
Eq.~(\ref{eq:11}) (solid lines) are plotted to evaluate the
experimental data ($\bullet$) obtained from Ref.~\cite{pablo33}.}   
\label{fig:3}
\end{center}
\end{figure}

Equation~(\ref{eq:11}) accommodates the unit for
$\rho(E_I^{y,x})$, which is $\Omega$ m$^{-1}$ (because the unit for
1D $n$ from Eq.~(\ref{eq:3}) is m$^{-1}$). The length, $l$ varies exponentially
as a result of Eq.~(\ref{eq:4}). Figure~\ref{fig:3}
a) and b) indicate the influence of length on resistance via
Eq.~(\ref{eq:11}). The $\bullet$ in Fig.~\ref{fig:3} a) and b)
represent the experimental data from de Pablo {\it et al.} for the
nanotube samples with diameters, $d$ = 1.5 nm and 1.7 nm
respectively. The solid lines are based on Eq.~(\ref{eq:11}).

Importantly, the fittings in Fig.~\ref{fig:3} a) and b) clearly
demonstrate that Eq.~(\ref{eq:11}) gives a reasonable
approximation. With this model at our disposal, one can utilize
the fitting parameters namely, $\pi dA$(13062) = 37 k$\Omega$ for
$d$ = 1.5 nm whereas $\pi dA$(13062) = 300 k$\Omega$ for $d$ = 1.7
nm. Therefore, $A_{1.5}$ = 6.01 $\times$ 10$^{8}$
s$^{-1}$K$^{-2}$ and $A_{1.7}$ = 4.30 $\times$ 10$^{9}$
s$^{-1}$K$^{-2}$. As a result of this, the $e$-$e$ scattering rate
for 1.5 nm and 1.7 nm nanotubes are respectively given by
$\tau_{e-e}$ = 1.85 $\times$ 10$^{-14}$ s and $\tau_{e-e}$ =
2.58 $\times$ 10$^{-15}$ s. Eventually, the mean free path, $l_e
= \upsilon_F \times \tau_{e-e}$ = (8.1 $\times$ 10$^{5}$)(1.85
$\times$ 10$^{-14}$) = 15 nm for $d$ = 1.5 nm, and for $d$ = 1.7
nm, $l_e$ = 2 nm. Here, the Fermi velocity, $\upsilon_F$ is
obtained from Ref.~\cite{kociak5}. The other fitting parameter,
$B$ for $d$ = 1.5 nm and 1.7 nm are found to be 0.47 and 0.18
respectively. Throughout this resistance calculations, $e$-$ph$
scattering has been neglected in the usual sense, because iFDS
have had the electrons dressed with $E_I$. Meaning, the excitation
of electrons and holes varies with different types of atoms (in
this case \texttt{C}), identically with the traditional methods
discussed by Barnett {\it et al.}~\cite{barnett36}, Perebeinos
{\it et al.}~\cite{perebeinos37} and Ando~\cite{ando46}. Contrary
to iFDS, the latter methods utilize the free-electron theory and
subsequently the $e$-$ph$ interaction was determined in order to
couple it with those free-electrons so as to describe the
excitation of electrons and holes with different types of atoms.
Consequently, one can notice that Eq.~(\ref{eq:11}) does not
ignore $e$-$ph$ interactions in any way. In fact, the existence of
polaronic effect via $E_I$ has been discussed using
iFDS~\cite{andrew27}. Parallel to this, Perebeinos {\it et
al.}~\cite{perebeinos37} have also found strong polaronic effect
in SNTs as inevitable.

The properties of phonons and its influence in CNTs specifically
and other nanostructures generally have been discussed extensively
in the
Refs.~\cite{xia38,ando39,pipinys40,maeda41,dutta42,jorio43,bose44,mensah47}.
Apart from that, Chen {\it et al.}~\cite{chen48} pointed out the
possibility of superconductivity and ferromagnetism in SNTs doped
by a chain of \texttt{C} atoms. Whereas, Ichida {\it et
al.}~\cite{ichida45} have carried out the necessary analysis on
the relaxation dynamics of photoexcited states in SNTs using
femtosecond spectroscopy. They found an interesting relationship
of which, the $e$-$ph$ interaction increases with decreasing tube
diameter. Qualitatively, their result explains why for small $d$
(1.5 nm), the $B$ (0.47) determined earlier is 2.6$\times$ larger than the
magnitude of $B$ (0.18), which is for large $d$ (1.7 nm). Recall here that $B$
corresponds to $E_I$, which is associated to the heavier effective mass (polaronic effect). In other words, this polaronic effect is due to
the interaction between non free-electrons and phonons, which
enhances the effective mass of the charge
carriers~\cite{andrew27}. On the contrary, for the well known $e$-$ph$
interaction in metals, free-electrons and phonons interact, that
eventually gives rise to $e$-$ph$ scattering. Having said that, we can now compare our predicted values for $l_e$ (2 to 15 nm) with the values obtained by considering the short optical phonon mean-free-path (10 to 20 nm, for low bias-voltage and $d$ = 1.5 to 2 nm) that limits the electrons mean-free-path~\cite{vasili}.    

\section{Conclusions}

In conclusion, the ionization energy based Fermi-Dirac statistics
has been employed to derive the length-dependent resistance in a
single-wall Carbon nanotube. It has been shown that such
dependence is inevitable in a low dimensional and
non-free-electron systems at nanoscales by using the the recent
experimental findings. In this paper, it is also highlighted that
simple equations derived using iFDS are able to capture the
transport properties of single-wall Carbon nanotubes with
reasonable accuracy.

\section*{Acknowledgments}

The authors would like to thank the School of Physics, University of Sydney and Professor Catherine Stampfl for the research opportunities.

\end{document}